\documentclass{aa}  

\usepackage{graphicx}
\usepackage{color}
\usepackage{txfonts}

\newcommand{\ha}{H$\alpha$}
\newcommand{\hb}{H$\beta$}
\newcommand{\oiii}{[O III]}
\newcommand{\nii}{[N II]}
\newcommand{\oii}{[O II]}
\newcommand{\micron}{\rm \mu m}

\begin{document}

   \title{The NIRSpec Wide GTO Survey}
   %\subtitle{I. Survey description }
   \author{Michael V. Maseda\inst{\ref{inst1}}
   \and Anna de Graaff\inst{\ref{inst2}}
   \and Marijn Franx\inst{\ref{inst3}}
   \and Hans-Walter Rix\inst{\ref{inst2}}
   \and Stefano Carniani\inst{\ref{inst4}}
   \and Isaac Laseter\inst{\ref{inst1}}
   \and Ugn\.{e} Dudzevi\v{c}i\={u}t\.{e}\inst{\ref{inst2}}
   \and Tim Rawle\inst{\ref{inst5}}
   \and Eleonora Parlanti\inst{\ref{inst4}}
   \and Santiago Arribas\inst{\ref{inst6}}
   \and Andrew J. Bunker\inst{\ref{inst7}}
   \and Alex J. Cameron\inst{\ref{inst7}}
   \and Stephane Charlot\inst{\ref{inst8}}
   \and Mirko Curti\inst{\ref{inst13}}
   \and Francesco D'Eugenio\inst{\ref{inst9},\ref{inst10}}
   \and Gareth C. Jones\inst{\ref{inst7}}
   \and Nimisha Kumari\inst{\ref{inst14}}
   \and Roberto Maiolino\inst{\ref{inst9},\ref{inst10}}
   \and Hannah \"Ubler\inst{\ref{inst9},\ref{inst10}}
   \and Aayush Saxena\inst{\ref{inst7}}
   \and Renske Smit\inst{\ref{inst11}}
   \and Chris Willott\inst{\ref{inst12}}
   \and Joris Witstok\inst{\ref{inst9},\ref{inst10}}
}
   
   \institute{Department of Astronomy, University of Wisconsin-Madison, 475 N. Charter St., Madison, WI 53706, USA\label{inst1}
   \and Max-Planck-Institut f\"ur Astronomie, K\"onigstuhl 17, 69117 Heidelberg,  Germany\label{inst2}   
   \and Leiden Observatory, Leiden University, P.O. Box 9513, 2300RA Leiden, the Netherlands\label{inst3}
      \and Scuola Normale Superiore, Piazza dei Cavalieri 7, I-56126 Pisa, Italy\label{inst4}
      \and European Space Agency (ESA), European Space Astronomy Centre (ESAC), Camino Bajo del Castillo s/n, 28692 Villafranca del Castillo, Madrid, Spain\label{inst5}
      \and Centro de Astrobiolog\'ia (CAB), CSIC–INTA, Cra. de Ajalvir Km.~4, 28850- Torrej\'on de Ardoz, Madrid, Spain\label{inst6}
      \and Department of Physics, University of Oxford, Denys Wilkinson Building, Keble Road, Oxford OX1 3RH, UK\label{inst7}
      \and Sorbonne Universit\'e, CNRS, UMR 7095, Institut d'Astrophysique de Paris, 98 bis bd Arago, 75014 Paris, France\label{inst8}  
      \and European Southern Observatory, Karl-Schwarzschild-Strasse 2, 85748 Garching bei M\"unchen, Germany\label{inst13}
      \and Kavli Institute for Cosmology, University of Cambridge, Madingley Road, Cambridge, CB3 0HA, UK\label{inst9}
      \and Cavendish Laboratory, University of Cambridge, 19 JJ Thomson Avenue, Cambridge, CB3 0HE, UK\label{inst10}
      \and AURA for European Space Agency (ESA), ESA Office, Space Telescope Science Institute, 3700 San Matin Drive, Baltimore, MD, 21218, USA\label{inst14}
      \and Astrophysics Research Institute, Liverpool John Moores University, 146 Brownlow Hill, Liverpool L3 5RF, UK\label{inst11}
      \and NRC Herzberg, 5071 West Saanich Rd, Victoria, BC V9E 2E7, Canada\label{inst12}
      }

   %\date{Received September 15, 1996; accepted March 16, 1997}

\titlerunning{}
\authorrunning{Maseda et al.}

  \abstract{The Near-infrared Spectrograph (NIRSpec) on the James Webb Space Telescope is uniquely suited to studying galaxies in the distant Universe with its combination of multi-object capabilities and sensitivity over a large range in wavelength ($0.6-5.3\,\micron$). Here we present the NIRSpec Wide survey, part of the NIRSpec Instrument Science Team's Guaranteed Time Observations, using NIRSpec's microshutter array to obtain spectra of more than 3200 galaxies at $z>1$ at both low- and high-resolution ($R\approx100$ and 2700) for a total of 105 hours.  With 31 pointings covering $\approx$320 arcmin$^2$ across the five CANDELS fields with exquisite ancillary photometry from the Hubble Space Telescope, the NIRSpec Wide survey represents a fast and efficient way of probing galaxies in the early Universe.  Pointing centers are determined to maximize the observability of the rarest, high-value sources. Subsequently, the microshutter configurations are optimized to observe the maximum number of ``census'' galaxies with a selection function based primarily on HST/F160W magnitude, photometric/slitless grism redshift, and predicted \ha\ flux tracing the bulk of the galaxy population at cosmic noon ($z_{\rm med}=2.0$).  We present details on the survey strategy, the target selection, an outline of the motivating science cases, and discuss upcoming public data releases to the community.
  }


   \maketitle

\section{Introduction}

The James Webb Space Telescope \cite[JWST;][]{2023PASP..135f8001G} represents a substantial leap forward in our ability to study galaxies in the distant Universe.  Its wavelength coverage beyond 2.4 $\mu$m, the effective limit for ground-based spectroscopy of faint targets, permits studies of the restframe-optical region of galaxy spectra at and beyond ``cosmic noon'' at $1 < z < 3$.  The Near-infrared Spectrograph \cite[NIRSpec;][]{2022AA...661A..80J} in particular offers spectral coverage from 0.6 to 5.3 $\mu$m at a variety of spectral resolutions for up to $\approx$200 galaxies simultaneously using the micro-shutter array \cite[MSA;][]{2022AA...661A..81F}.  

\begin{figure}
\begin{center}
\includegraphics[width=0.49\textwidth]{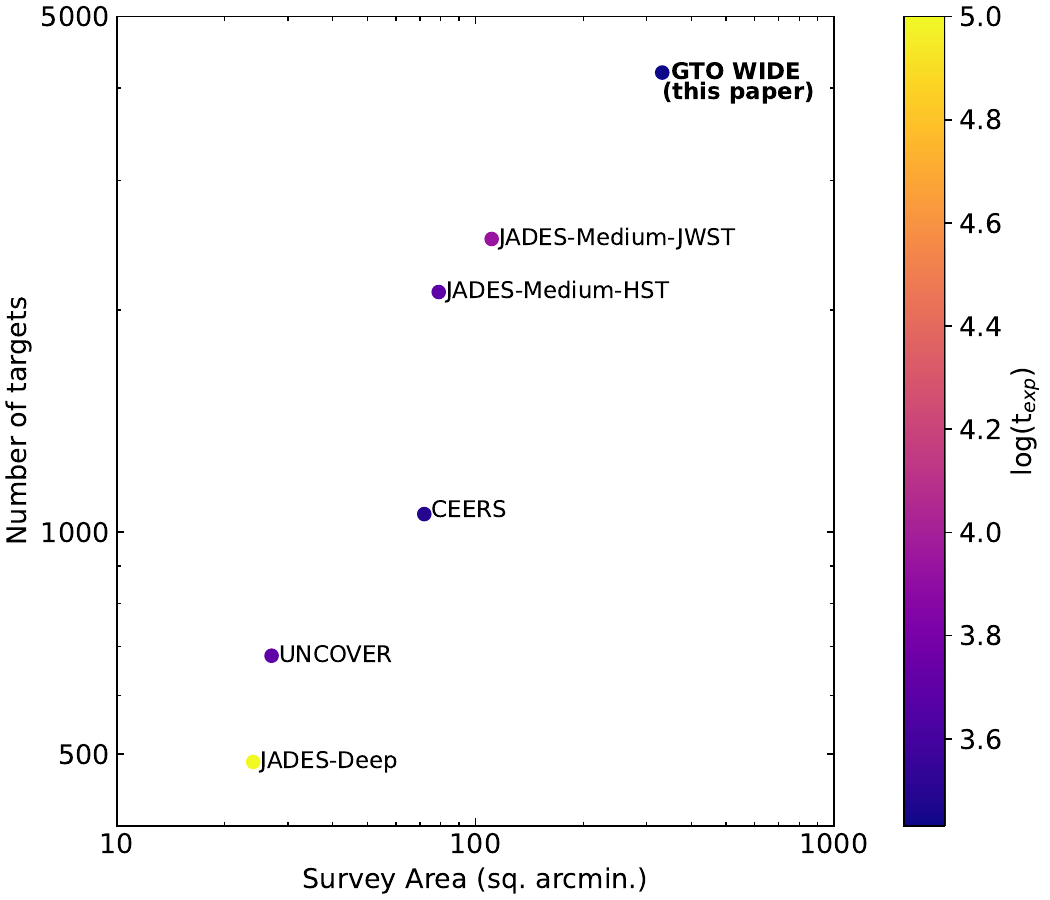}
\end{center}
\caption{Area covered by the NIRSpec MSA versus total number of targets observed in the survey for various Cycle 1 programs (limited to PRISM observations).  While amongst the shallowest in terms of (PRISM) exposure time, the NIRSpec ``Wide'' GTO survey covers the largest area and targets the most individual galaxies.  In addition, the Wide survey is spread over five non-contiguous fields in order to maximize the chances of observing cosmologically-rare targets.}
\label{fig:survey}
  \end{figure}

To harness the capabilities of NIRSpec, a multi-tiered Guaranteed Time Observations (GTO) program was conceived with a primary focus on the growth and evolution of galaxies across cosmic time.  Multiple tiers can serve as a way to mitigate the traditional trade-off between survey depth and survey area.  The ``deep'' and ``medium'' tiers of the extragalactic multi-object portion of the NIRSpec GTO program are combined with deep imaging from the NIRCam instrument, with the collective joint program referred to as the JWST Advanced Deep Extragalactic Survey \cite[JADES;][]{2023arXiv230602465E}.  However, even the broadest-area spectroscopic coverage in JADES (at a ``medium'' depth of $\approx$ 29 magnitude) covers less than 175 square arcminutes in total in two extragalactic fields \citep[GOODS-N and GOODS-S;][]{Giavalisco:2004}, as shown in Figure \ref{fig:survey}.

At a discovery-space level, the motivation for an even wider tier is driven by the fact that even relatively short exposures with NIRSpec can obtain data of a depth and wavelength coverage completely unattainable with other facilities \cite[e.g.][]{2023ApJ...951L..22A,2023ApJ...949L..25F}.  This is further enhanced with the unique multi-object spectroscopic (MOS) capabilities of NIRSpec, allowing for powerful large-scale surveys of faint targets.  The NIRSpec Wide GTO Program (henceforth referred to as Wide), allocated 105 hours, is the fastest general covering of the CANDELS fields \citep{Grogin:2011,Koekemoer:2011} possible with NIRSpec while still maintaining an overall science-to-total efficiency of $>$ 50 \% when accounting for instrument/facility overheads.  Wide is less than a magnitude shallower but covers an area twice as large than the ``JWST-Medium'' layer in JADES.

Instead of multiple spectroscopic visits to the same area of the sky to maximize completeness \cite[e.g. the strategy of the JADES Deep and UNCOVER surveys;][]{2023arXiv230602465E,2022arXiv221204026B}, Wide maximizes the independent area covered by NIRSpec to characterize the rarest galaxy populations with source densities on the order of 1 per NIRSpec field-of-view (FoV excluding gaps $\approx$ 9 arcmin$^2$) or less.  To go as wide as possible in Cycle 1, we utilize the existing larger-area photometric survey CANDELS, performed with the \textit{Hubble} Space Telescope (HST).  In particular, we focus on the $\approx$ 900 square arcminutes with homogeneously produced photometric and WFC3/G141 slitless grism spectroscopic catalogs from the 3D-HST survey \citep{2012ApJS..200...13B, 2014ApJS..214...24S, 2016ApJS..225...27M}.  The photometric data spans from the UV to the near-IR, including multi-band HST and \textit{Spitzer}/IRAC photometry.  The extensive filter coverage in these fields combined with imaging sensitivities that are well-matched to our intended (short) exposure times make these HST legacy fields ideally-suited for target selection in Wide.

The science cases facilitated by a wide-area spectroscopic survey covering 0.6 to 5.3\,$\mu$m are myriad.  Focusing on the intermediate- to high-$z$ Universe ($z \gtrsim 1$), a survey like Wide is ideally suited to understanding the nature of galaxy growth and quenching across cosmic time, including the peak of the cosmic star formation rate density at $1 < z < 2$ \citep{MadauDickinson}.  Even spectra of modest resolution ($R\approx30-300$) can provide significantly more detailed information about the stellar and ionized gas properties in distant galaxies than photometry alone \cite[e.g.][]{2010ApJ...718L..73V,2023NatAs...7..622C}.  In addition, high-resolution spectroscopy (here, $R\approx2700$) can be a key diagnostic in understanding the kinematic properties of the ionized gas in galaxies, including measurements of dynamical masses \citep{deGraaff2023} or the interplay between the host galaxy and an accreting supermassive black hole.  By combining these two spectral modes for a sample of more than 3200 galaxies at $z > 1$, the Wide survey will provide a key reference point and treasury dataset for future NIRSpec surveys focused on galaxy growth and evolution across cosmic time.

In this paper, we present the Wide Survey with a characterization of the observation strategy (Section \ref{sec:obs}), the target selection (Section \ref{sec:targets}), the data reduction (Section \ref{sec:red}), the primary science cases (Section \ref{sec:science}), and a data release plan (Section \ref{sec:release}). Throughout we assume a flat $\Lambda$CDM cosmology with $H_0$ = 70 km s$^{-1}$ Mpc$^{-1}$ and $\Omega_m=0.3$, and we utilize AB magnitudes \citep{1983ApJ...266..713O}.

%Figure ideas:
%\begin{enumerate}
    %\item HST images with NIRSpec footprints (also label JADES, PRIMER, ... ?)
    %\item Sampling in M*-SFR-z space
%    \item Representative spectra and expected vs. delivered SNR
    %\item Initial version of zspec-zphot plot
%    \item Initial comparison of UV+IR SFR versus Balmer lines (plus dust)
    %\item UVJ with Halpha EWs/limits
    %\item R2700 outflow example (comparison to IFS?)
%\end{enumerate}

\section{Observational Summary}
\label{sec:obs}
The major science goals (described in Section \ref{sec:science}) set the basic requirements for the NIRSpec observations. Low-resolution spectra of moderate depth are sufficient to understand the general stellar and ionized gas properties of galaxies over a wide redshift range. At the same time, we need high-resolution spectra to accurately characterize the galaxy kinematics and galactic outflows, and to resolve closely-spaced emission lines such as \nii\ $\lambda\lambda$6548, 6584 and \ha\ for use in e.g. active galactic nuclei (AGN) diagnostics. As described below, we designed our MSA configurations to allow us to simultaneously obtain spectra of $\gtrsim$130 galaxies. For the low-resolution spectra, we use the PRISM disperser with the CLEAR filter which covers a wavelength range of 0.6$\mu$m--5.3$\mu$m at a spectral resolution of $R\approx100$ \cite[varying from 30 to 300 for a uniformly illuminated shutter;][]{2022AA...661A..80J}. The high resolution spectra are obtained with the G235H and G395H gratings (and associated F170LP and F290LP filters), providing a maximum wavelength coverage of 1.66$\mu$m--3.05$\mu$m and 2.87$\mu$m--5.14$\mu$m, respectively, at a spectral resolution of $R\approx$ 2700. Using two\footnote{For a single pointing, 09 in GOODS-S, we include an additional observation in G140M/F090LP ($R\approx$ 1000).  This is designed to assess the utility of shorter-wavelength/lower-resolution data to several of our priority science cases.} high resolution gratings allows us to obtain multiple important optical diagnostics lines, such as \oiii\ $\lambda5007$, \ha, [N II], and [S II] $\lambda\lambda6717,6730$, at sufficient spectral resolution to resolve multiple kinematic components and dis-entangle closely spaced lines for a given galaxy over a wide redshift range from 2.3 to 6.6.% that is inaccessible with any other instrument. 

The MSA features almost 250,000 micro shutters with a projected open size of $\sim0\farcs20\times0\farcs46$ which are structured in four nearly squared quadrants assembled of 365 shutters in the dispersion direction and 171 shutters in the cross-dispersion direction \citep{2022AA...661A..81F}. The full patrol field of the MSA is about $3.6\arcmin\times3.4\arcmin$ with small gaps in between the quadrants. About 15\% of shutters are unable to open \citep{2022SPIE12180E..3RR,2023PASP..135c8001B} which adds complexity in the mask design. For Wide we used the standard scheme of opening 3 shutters per target aligned in the cross-dispersion direction to create a pseudo-slit of $0\farcs2\times1\farcs5$ to facilitate sky subtraction via nodding. % for the majority of sources. 

Using the high-resolution modes requires careful planning as the dispersed spectra cannot always be fully covered by the detector array, depending on the position of the shutter \cite[see also][]{2019MNRAS.486.3290M, 2022AA...661A..80J, 2022AA...661A..81F}. As such, objects lying in different parts of the MSA area will have different spectral coverage due to truncation, in addition to the obscuration by the gap between the NRS1 and NRS2 detectors.  Another limitation is that the maximum number of targets that can be observed without overlapping spectra in the high-resolution setups is about a factor of 4 smaller than for the PRISM setup. 

For Wide we define a strategy to mitigate these limitations and maximize the observable number of targets. The impact of the limited wavelength coverage of the high-resolution setups can be mitigated by enforcing a redshift-dependent constraint for assigning targets to specific shutters so that specific emission lines are still covered on the detector. This process will be described in detail in Section \ref{sec:pclass}. Since the expected count rates per pixel in the high-resolution modes will be small for the zodiacal background and the continuum emission of the galaxies, we allowed these spectra to overlap on the detector. As the emission lines will occupy only a very small area on the detector, the chances for overlapping emission lines will be minimal whereas any background can be subtracted locally (see Section \ref{sec:red}).  Therefore, we used the same MSA configuration in PRISM and G235H/G395H, allowing spectra to overlap in the high-resolution modes.

\subsection{Survey fields}
Given the high angular resolution of JWST and the small ``slit'' size of the MSA, it is necessary to know target coordinates to better than 25 milliarcsec. %This clearly demands survey fields with existing deep HST imaging. Therefore, 
The Wide survey is based on the Cosmic Assembly Near-infrared Deep Extragalactic Legacy Survey \citep[CANDELS;][]{Grogin:2011,Koekemoer:2011}, which provides high-quality NIR imaging for five extensively studied regions on the sky from the two Great Observatories Origins Deep Survey \citep[GOODS,][]{Giavalisco:2004} fields north and south, i.e. GOODS-N and GOODS-S, the Cosmic Evolution survey \citep[COSMOS,][]{Scoville:2007},  the All-wavelength Extended Groth Strip International Survey \citep[AEGIS,][]{Davis:2007}, and the UKIDSS Ultra-deep Survey field \citep[UDS,][]{Lawrence:2007}. In all five of these fields, we use astrometry that is tied to the Gaia DR2 reference frame \citep{2012AA...538A..78L}, based on re-reduced HST mosaics (G. Brammer: \url{https://s3.amazonaws.com/grizli-stsci/Mosaics/index.html})

\begin{figure*}
 
\begin{center}
\includegraphics[width=0.95\textwidth]{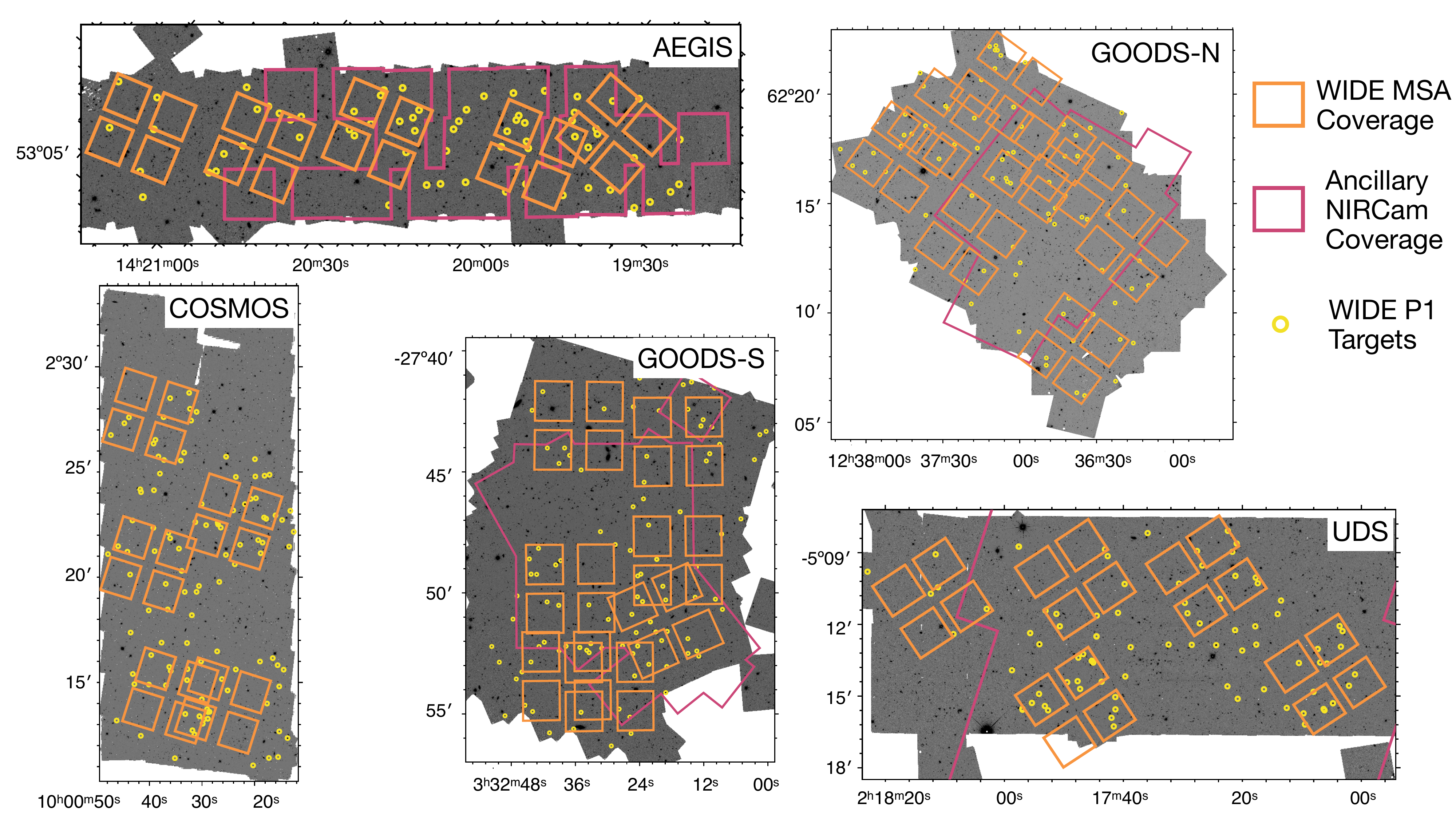}
\end{center}
\caption{The NIRSpec Wide pointings within the CANDELS fields.  The background images are the F160W mosaics from all HST programs in these fields (G. Brammer, private communication).  Overlaid are the NIRCam imaging regions from CEERS \cite[AEGIS;][]{2023ApJ...946L..12B}, PRIMER (UDS; PI: J. Dunlop; JWST-GO-1837), and JADES \cite[GOODS-N and GOODS-S;][]{2023arXiv230602465E}; COSMOS-Web \citep{2023ApJ...954...31C} and PRIMER-COSMOS cover the full area of the COSMOS field shown here.}
  \label{fig:pointings}
  \end{figure*}

Due to the orientation requirements of the JWST spacecraft, each of the five fields can only be observed during specific observing windows with a limited range of position angles. To aid flexibility in the observation scheduling, the NIRSpec Wide survey did not request fixed position angles to control the NIRSpec MSA orientation beforehand. %Hence, the exact orientation of the NIRSpec MSA remains unknown until actual scheduling.  
The general pointing centers are chosen to maximize the potential observability of the highest-priority sources (Section \ref{sec:p1}), using a patrol field of 9\arcmin\ diameter \citep{overbooking}. The final pointing centers as well as the assigned NIRSpec position angles (APA, not V3PA) are listed in Table~\ref{tab:Field_coors}. In total, Wide consists of 31 NIRSpec MSA pointings of which 5 are in AEGIS, 5 are in COSMOS, 5 are in UDS, 9 are in GOODS-N, and 7 are in GOODS-S: see Figure \ref{fig:pointings}.

\begin{table}
\begin{center}
 \begin{tabular}{ccccc}\hline\hline
  Field & RA & Dec & APA & Obs. date\\
     & (J2000) & (J2000) & (deg) & \\\hline

  \multicolumn{5}{c}{GOODS-N (PID 1211)}\\\hline
  00 &   12:36:16  &  62:13:05 & 318.70 & 2023-03-26 \\
  01 &   12:36:38  &  62:16:28 & 322.76 & 2023-03-25 \\
  02 &   12:36:39  &  62:08:23 & 322.57 & 2023-03-25 \\
  03 &   12:37:06  &  62:20:20 & 322.90 & 2023-03-28 \\
  04 &   12:37:20  &  62:13:22 & 322.77 & 2024-03-24*\\
  05 &   12:37:22  &  62:16:07 & 31.21 & 2024-01-30* \\
  06 &   12:37:47  &  62:17:08 & 322.94 & 2023-03-26 \\
  07 &   12:36:53  &  62:16:40 & 322.79 & 2023-03-26 \\
  08 &   12:37:30  &  62:18:38 & 322.96 & 2023-03-26 \\\hline
  \multicolumn{5}{c}{GOODS-S (PID 1212)}\\\hline
  09 &  03:32:16  & -27:48:02 & 89.52 & 2023-10-14 \\ 
  10 &  03:32:36  & -27:42:02 & 88.52 & 2023-10-14 \\
  11 &  03:32:12  & -27:43:59 & 89.49 & 2023-10-08 \\
  12 &  03:32:18  & -27:51:30 & 202.74 & 2024-01-27*\\
  13 &  03:32:33  & -27:50:15 & 89.56 & 2023-10-12 \\
  14 &  03:32:28  & -27:54:09 & 89.52 & 2023-10-13 \\
  15 &  03:32:42  & -27:53:14 & 89.55 & 2023-10-14 \\\hline
  \multicolumn{5}{c}{AEGIS (PID 1213)}\\\hline
  17 &  14:20:38  &  53:03:42 & 17.34 & 2023-03-08 \\
  18 &  14:20:20  &  52:59:16 & 17.24 & 2023-03-08 \\
  19 &  14:19:55  &  52:55:49 & 17.11 & 2023-03-09 \\
  20 &  14:19:31  &  52:50:09 & 17.01 & 2023-03-09 \\
  21 &  14:19:11  &  52:48:00 & 268.75 & 2023-06-28*\\\hline  
  \multicolumn{5}{c}{COSMOS (PID 1214)}\\\hline
  23 &  10:00:40  &  02:27:39 & 71.56 & 2023-12-02\\  
  24 &  10:00:24  &  02:22:38 & 71.55 & 2023-12-08\\  
  25 &  10:00:40  &  02:20:36 & 71.56 & 2023-12-09\\  
  27 &  10:00:36  &  02:14:22 & 71.56 & 2024-01-06\\  
  28 &  10:00:26  &  02:13:53 & 71.55 & 2023-12-03\\   \hline
  \multicolumn{5}{c}{UDS (PID 1215)}\\\hline
  29 &  02:18:12  & -05:10:55 & 32.47 & 2023-07-23 \\   
  30 &  02:17:48  & -05:10:08 & 32.45 & 2023-08-19 \\   
  31 &  02:17:48  & -05:15:29 & 32.43 & 2023-08-19 \\  
  32 &  02:17:26  & -05:09:59 & 32.43 & 2023-08-19 \\  
  33 &  02:17:06  & -05:14:06 & 32.41 & 2023-08-20\\  \hline
 \end{tabular}
\caption{Observed/Planned Wide pointings. APA refers to the aperture position angle (in degrees), i.e. the angle that the rectangular microshutters make as projected onto the sky, not the ``V3PA'' of the instrument itself. Fields 16, 22, and 26 were initially planned but never observed and are omitted from this table. (*) observation was repeated as original exposures were affected by either a telescope guide star acquisition error or a bright short in the MSA control electronics}
\label{tab:Field_coors}
\end{center}
\end{table}

\subsection{Exposure times}
The aim of the Wide survey is to efficiently cover a large area while incurring minimal observing overheads. A fast general covering of the sky becomes efficient if the observatory and instrument overheads are less than $\approx$ 50\% of the total observing time. This sets a minimum of approximately 100 minutes per pointing of integration time when including the large telescope slew time to include each of the five CANDELS fields in the survey.  With a total time allocation of 105\,h for Wide from the overall NIRSpec GTO budget this permits us to observe 31 pointings, i.e. the scope of the Wide survey is set by observatory overheads\footnote{For this reason we opted not to include coordinated parallel observations, as they would incur telescope overheads.  However, given the prime location of the Wide pointings, pure parallel programs such as PANORAMIC (JWST-GO-2514; PIs C. Williams and P. Oesch) can obtain scientifically-useful data with no additional cost.}, naturally dove-tailing with our scientific priority of maximal non-contiguous areal coverage.

\begin{figure}
\includegraphics[width=0.45\textwidth]{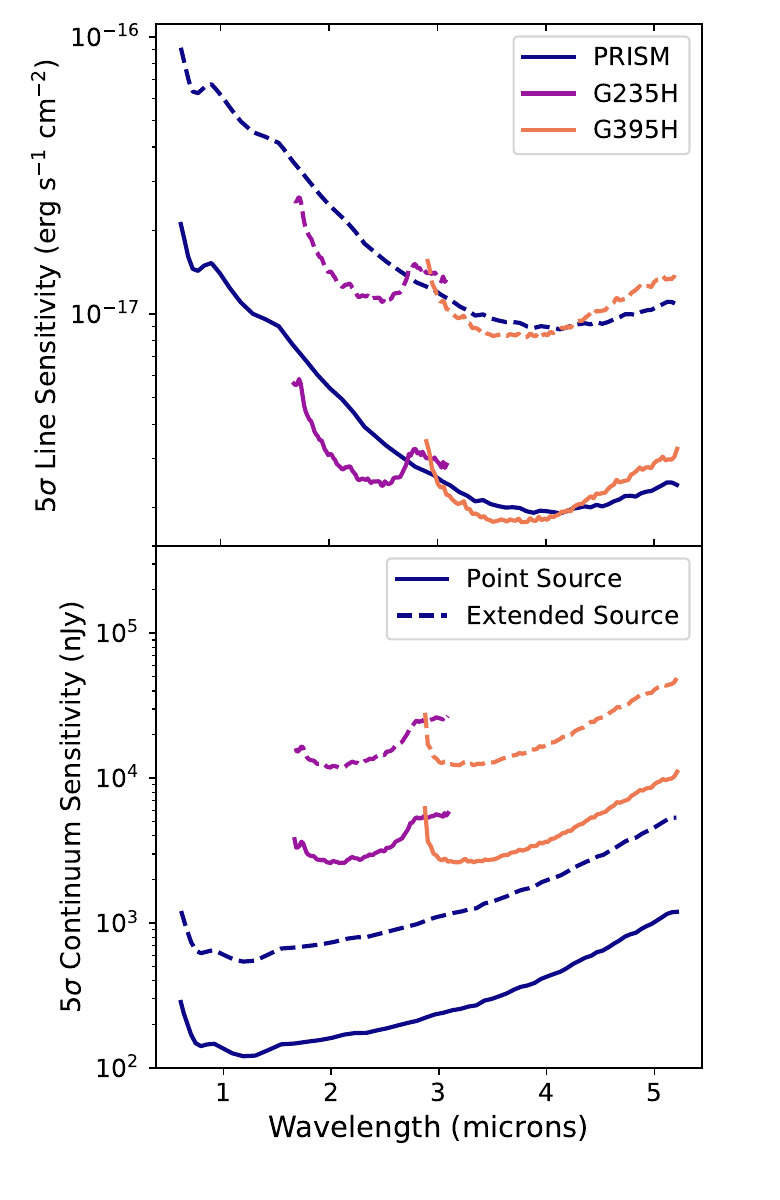}
\caption{Empirically-derived emission line (top) and continuum (bottom) sensitivities for the Wide survey as a function of disperser and observed wavelength.  The solid lines are the 5-$\sigma$ sensitivities for a centered point source, while the dashed lines are the 5-$\sigma$ sensitivities for an ``extended'' source, namely a centered object with a Sersic index $n=1$ and a half-light radius of 0$\farcs$3 \cite[a typical star-forming galaxy at $z\approx$1.5;][]{Wel:2014}.}
\label{fig:sens}
\end{figure}

Adopting the NRSIRS2RAPID read-out mode of the detectors \citep{2017PASP..129j5003R} and a standard 3-point nodding pattern to cycle through the 3 shutter-slitlet per target, we took 1 exposure with 55 groups for the PRISM spectral setup. This leads to an effective exposure time of 2451s (41 minutes). Due to the extended nature of the $z=1-3$ Wide galaxies, for both high-resolution gratings we only nodded between the two outer shutter positions to minimize source self-subtraction, i.e. the source is only ever centered in the upper and lower open shutter.  For G235H we took 1 exposure with 55 groups for each of the two nodding positions (1634s/27m total), and for G395H we took 1 exposure with 60 groups for each of the two nodding positions (1780s/30m total).  The small difference between G395H and G235H is motivated by obtaining the highest possible signal-to-noise in \ha\ emission line at the highest redshifts.  The associated sensitivities are shown in Figure \ref{fig:sens} \cite[see also][]{2022SPIE12180E..0XG}.

%\begin{table}
% \begin{tabular}{cccccc}\hline\hline
%  Disperser    & nods & $t_\mathrm{exp}$  & $f_\mathrm{line}$ & $f_\mathrm{cont}$ \\\hline
%  PRISM/CLEAR  & 3    & 41\,min           &	$1.5\times10^{-18}$   & 26.7\,mag \\
%  G235H/F170LP & 2    & 27\,min           &	$2\times10^{-18}$   & 23.1\,mag \\
%  G395H/F290LP & 2    & 30\,min           &	$2\times10^{-18}$   & 23.2\,mag \\\hline
  
% \end{tabular}
% \caption{\label{tab:integrations}
%Integration times per spectral setup. 
% Line and continuum sensitivities are 3-$\sigma$. \textbf{These are dummy numbers - will %include a figure}}
% \end{table}

\section{Target Selection and MSA Design}
\label{sec:targets}
Once a position angle was assigned to a pointing, we chose the optimal field center corresponding to the best observability of our highest-priority targets (Section \ref{sec:p1}).  The remaining MSA shutters were allocated to sources drawn from the 3D-HST catalog \citep{2016ApJS..225...27M}.  These ``census'' sources were prioritized according to a combination of their F160W magnitude (a direct observable), grism redshift \cite[which in some cases is simply a photometric redshift;][]{2012ApJS..200...13B,2018ApJ...854...29M}, UV plus IR star formation rate (derived assuming the grism redshift), and/or observability of optical emission lines in the G235H or G395H gratings.  We detail our priority classes in Table \ref{tab:pclass} and Section \ref{sec:pclass}.  These classes are used to optimize the final MSA configuration, the process of which is described in Section \ref{sec:empt}.

\subsection{Highest-priority Targets}
\label{sec:p1}
Individual high-priority objects were used to set the initial relative pointing centers during Phase I of the observing preparations, and then they were used again to define exact pointing centers once a final position angle was assigned (see Section \ref{sec:empt}), utilizing a weighting scheme for different classes.  These ``P1'' target categories are:
\begin{itemize}
    \item{Galaxies at $z > 1.5$ with estimated stellar masses $\gtrsim~10^{11.5}$ M$_{\odot}$ in the 3D-HST survey \cite[][weight = 1000]{2016ApJS..225...27M}.}
    \item{Sources that are continuum-bright in NIRCam F444W imaging (where available at the time of scheduling), but significantly fainter in HST F160W imaging. This selection is defined by both $\rm F444W<24.5$ and $\rm F160W > 25.2$, and is optimized to sample sources that would be missed by our default selection of continuum-bright sources based on F160W magnitude alone (weight = 500)}
    \item{IRAC-excess sources from \citet{Smit:2015,Roberts-Borsani:2016}, i.e. $z = 6-8$ sources with strong optical emission lines determined from \textit{Spitzer}/IRAC photometry (weight = 300).}
    \item{$z > 4$ photometric dropouts with $F160W < 24$ from \citeauthor{2015ApJ...803...34B} (\citeyear[][weight = 300]{2015ApJ...803...34B}).}
    \item{Compact quiescent galaxies at $z > 3$ from \cite[][weight = 300]{2016ApJ...830...51S}.}
    \item{Radio AGN ($L_{\mathrm{1.4GHz}} > 10^{24.5}$ W Hz$^{-1}$ from e.g. \citeauthor{2018MNRAS.473.2493M} (\citeyear{2018MNRAS.473.2493M}, weight = 120).}
    \item{X-ray AGN (hard X-ray luminosity $L > 10^{44.5}$ erg s$^{-1}$) from e.g. \citeauthor{2015MNRAS.451.1892A} (\citeyear{2015MNRAS.451.1892A}, weight = 60).}
    \item{Individual interesting sources such as GN-20 \citep{2009ApJ...694.1517D}, a bright submillimeter galaxy, or GN-z11 \citep{2016ApJ...819..129O,2023AA...677A..88B,2023arXiv230600953M},  the highest-redshift galaxy confirmed with HST (weight = 500).}
\end{itemize}

\begin{figure}
\begin{center}
\includegraphics[width=0.49\textwidth]{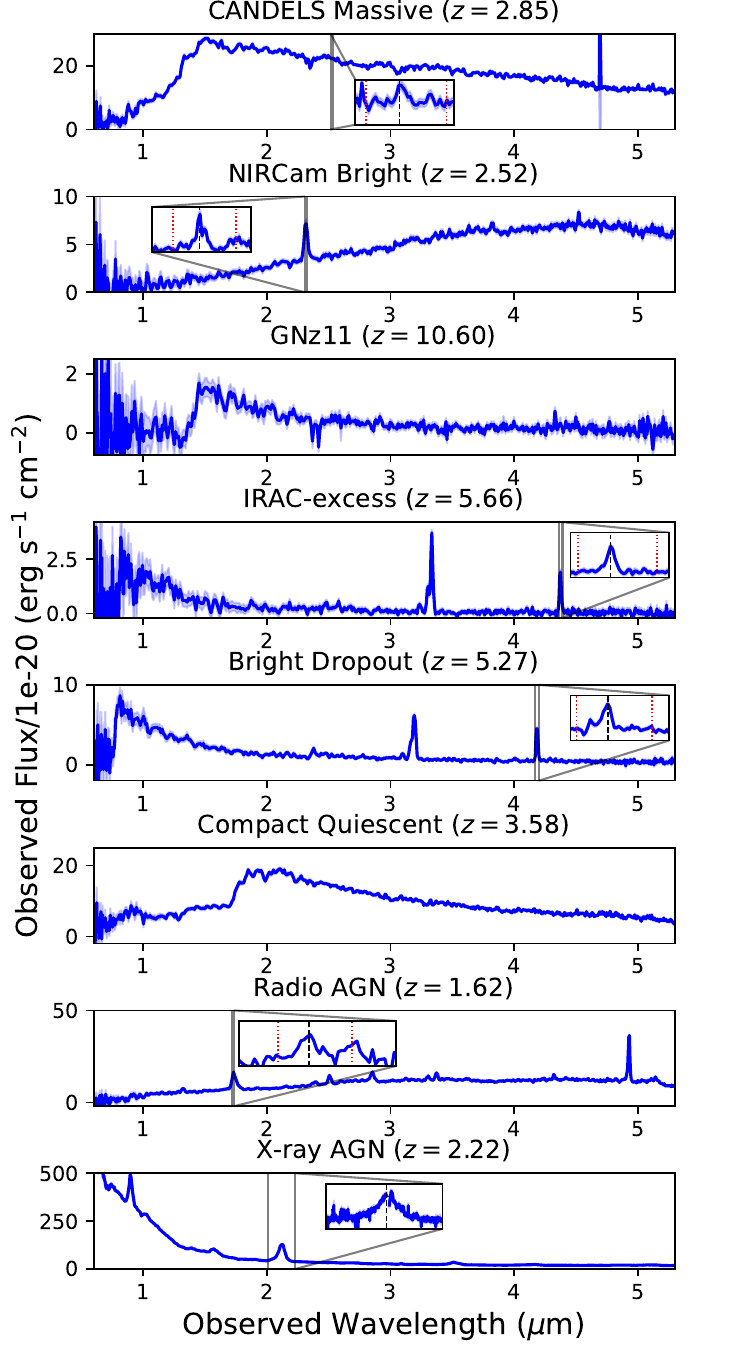}
\end{center}
\caption{Example PRISM spectra for objects in each of the highest-priority ``P1'' categories (Section \ref{sec:p1}), including the unique number observed in the survey when applicable.  Inset figures show high-resolution data covering \ha\ (vertical dashed lines) and \nii\ (vertical dotted lines), when detected.}
\label{fig:p1}
  \end{figure}

A representative object from each P1 class is shown in Figure \ref{fig:p1}.  In addition to these P1 AGN, we include an additional, lower-priority class of AGN (Priority Class 6) without the radio/X-ray luminosity cuts described above.  These samples are more heterogeneous, depending on the depth of the radio/X-ray data in the field, but nevertheless help increase the dynamic range of accretion rates and black hole masses probed in Wide.

\subsection{Priority Classes}
\label{sec:pclass}
One of the key motivators of the Wide survey with JWST/NIRSpec is to observe the restframe-optical part of a galaxy's spectrum at wavelengths that are difficult or impossible to observe with ground-based instruments.  As such, our highest priority classes primarily focus on the $z > 2.4$ regime where \ha\ becomes redshifted beyond the $K$-band.  As our high-resolution spectroscopy covers redwards of 1.66 $\mu$m, this means we also have high-resolution data for \oiii\ and \hb.  We therefore can cover the pre-eminent restframe-optical emission lines (\ha, \hb, and \oiii) out to $z \approx 7$, which includes the aforementioned \textit{Spitzer}/IRAC-excess sources.  At $1.5 < z < 2.4$ we do not have high-resolution coverage for \oiii\ and \hb, but we do typically have coverage of \ha: this represents our second-priority redshift range.  To summarize, our highest-priority redshift range is $z \ge 2.4$ where all strong optical lines are potentially visible in G235H or G395H (in practice, with the WFC3-based detection from 3D-HST our main census is limited to $z \lesssim 4-5$), followed by the range $1.5 \le z < 2.4$ where \ha\ is potentially visible in G235H, followed by objects at $z < 1.5$.  This tiered structure, with more objects in lower priority classes, is crucial to efficiently fill the MSA with targets.

One additional point is that the G235H and G395H gratings do not offer uniform wavelength coverage for all targets in the MSA field-of-view. The spectra are dispersed beyond the edge of the detector and/or are dispersed across the gap between the two detectors.  However, the precise wavelength coverage for every shutter is calculable. Combined with our photometric/grism redshifts we can, for a given spatial position, determine whether or not the specified emission lines are observable in one of the high-resolution gratings observed in Wide.

Thus, within these different redshift ranges we sub-prioritize based on the object's F160W magnitude, predicted \ha\ luminosity (based on the 3D-HST UV+IR star formation rates), and the possibility of observable optical emission lines in the high-resolution spectra.  This prioritization is performed \textit{after} an initial pointing position is generated (see Section \ref{sec:empt}) based on our highest priority targets, which we outline in the following section.  We note that the F160W selection prioritizes sources brighter than AB magnitude 24, which is the approximate limit where we expect our PRISM observations to produce a continuum signal-to-noise of 10 per resolution element for a source with a typical amount of geometrical slit losses ($\approx$1.4 magnitudes; see Figure \ref{fig:sens}).  Likewise, for predicted \ha\ fluxes above 10$^{-16.9}$ erg s$^{-1}$ cm$^{-2}$ we expect our high-resolution spectra to detect optical emission lines at sufficient signal-to-noise, in this case providing \hb\ at a signal-to-noise in excess of 3 at all wavelengths (modulo uncertainties on the star formation rates themselves and dust attenuation, and assuming \oiii\ and \ha\ will be visible at a signal-to-noise greater than 10).

\begin{table*}
 \begin{tabular}{cccccc}\hline\hline
  Priority    & Redshift & F160W  & H$\alpha$ Flux (erg s$^{-1}$ cm$^{-2}$)* & Other \\\hline
1 & -- & -- & -- & (see Section \ref{sec:p1})\\
2 & $z \ge 2.4$ & $F160W \le 24$ & $\mathrm{log}~ f_{\mathrm{H\alpha}} < -16.9$ & --  \\
3 & $1.5 \le z < 2.4$ & $F160W \le 24$ & $\mathrm{log}~ f_{\mathrm{H\alpha}} < -16.9$ & -- \\
4 & $z \ge 2.4$ & -- & $\mathrm{log}~ f_{\mathrm{H\alpha}} \ge -16.9$ & Coverage of \ha, \hb, and \oiii\ in G235H or G395H \\
5 & $z \ge 2.4$ & -- & $\mathrm{log}~ f_{\mathrm{H\alpha}} \ge -16.9$ & Coverage of \ha\ in G235H or G395H (cf. Priority class 4) \\
6 & -- & -- & -- & Selected as an AGN (see Section \ref{sec:p1}) \\
7 & $z \ge 2.4$ & $F160W > 24$ & $\mathrm{log}~ f_{\mathrm{H\alpha}} \ge -16.9$ & Coverage of \ha, \hb, and \oiii\ in G235H or G395H \\
8 & $z \ge 2.4$ & $F160W > 24$ & $\mathrm{log}~ f_{\mathrm{H\alpha}} \ge -16.9$ & Coverage of \ha\ in G235H or G395H (cf. Priority class 7) \\
9 & $1.5 \le z < 2.4$ & $F160W \le 24$ & $\mathrm{log}~ f_{\mathrm{H\alpha}} \ge -16.9$ & Coverage of \ha\ in G235H or G395H \\
10 & $1.5 \le z < 2.4$ & $F160W > 24$ & $\mathrm{log}~ f_{\mathrm{H\alpha}} \ge -16.9$ & Coverage of \ha\ in G235H or G395H \\
11 & $z \ge 2.4$ & $F160W > 24$ & $\mathrm{log}~ f_{\mathrm{H\alpha}} \ge -16.9$ & (cf. Priority class 7)  \\
12 & $1.5 \le z < 2.4$ & $F160W \le 24$ & $\mathrm{log}~ f_{\mathrm{H\alpha}} \ge -16.9$ & (cf. Priority class 9) \\
13 & $1.5 \le z < 2.4$ & $F160W > 24$ & $\mathrm{log}~ f_{\mathrm{H\alpha}} \ge -16.9$ & (cf. Priority class 10) \\
14 & $z < 1.5$ & $F160W \le 24$ & $\mathrm{log}~ f_{\mathrm{H\alpha}} < -16.9$ & (Satisfies one or both of the quoted magnitude/\ha\ cuts) \\
15 & $z < 1.5$ & $F160W > 22$ & -- & --  \\\hline
 \end{tabular}
 \caption{Wide Priority Classes, as used in the allocation of MSA slitlets (see Section \ref{sec:targets}).  For a source that fulfills multiple criteria, e.g. a source that could be in priority class 4 or 5, the source is put into the highest possible priority class (class 4, in this case).  Note that below a redshift of $z=2.4$ we no longer have coverage of \hb\ or \oiii\ in G235H. (*) Derived from the UV+IR star formation rate at $z_{\rm grism}$ \citep{2016ApJS..225...27M}, assuming Case B recombination and no dust attenuation.}
 \label{tab:pclass}
 \end{table*}

\subsection{MSA Configurations with eMPT}
\label{sec:empt}

In order to make MSA configurations, we used the above-mentioned priority classes as input to the eMPT software \cite[][]{eMPT}\footnote{\url{https://github.com/esdc-esac-esa-int/eMPT_v1}}.  The eMPT code operates in two stages.  First, the optimal pointing is determined by maximizing the simultaneous observability of ``P1'' or the highest-priority sources (Section \ref{sec:p1}) using the Initial Pointing Algorithm (IPA) incorporating the weights assigned to each P1 category.  This results in a set of pointings that maximize the observability of the highest-value P1 sources.  Second, at the best pointing centers from the IPA, a figure-of-merit is calculated based on the observability of sources in each of the subsequent priority classes (Table \ref{tab:pclass}).  The optimal pointing is then the one that maximizes the simultaneous observability of P1 targets as well as all of the other priority classes with decreasing importance.  Allowing for overlaps on the detector in the high-resolution modes but not the PRISM mode (Section \ref{sec:obs}) allowed us to observe a mean of 135 targets per configuration. This number ranges from 111 to 151 targets per pointing, scaling predominantly with the source density of targets in the parent catalog.

      \begin{figure*}
\begin{center}
\includegraphics[width=0.95\textwidth]{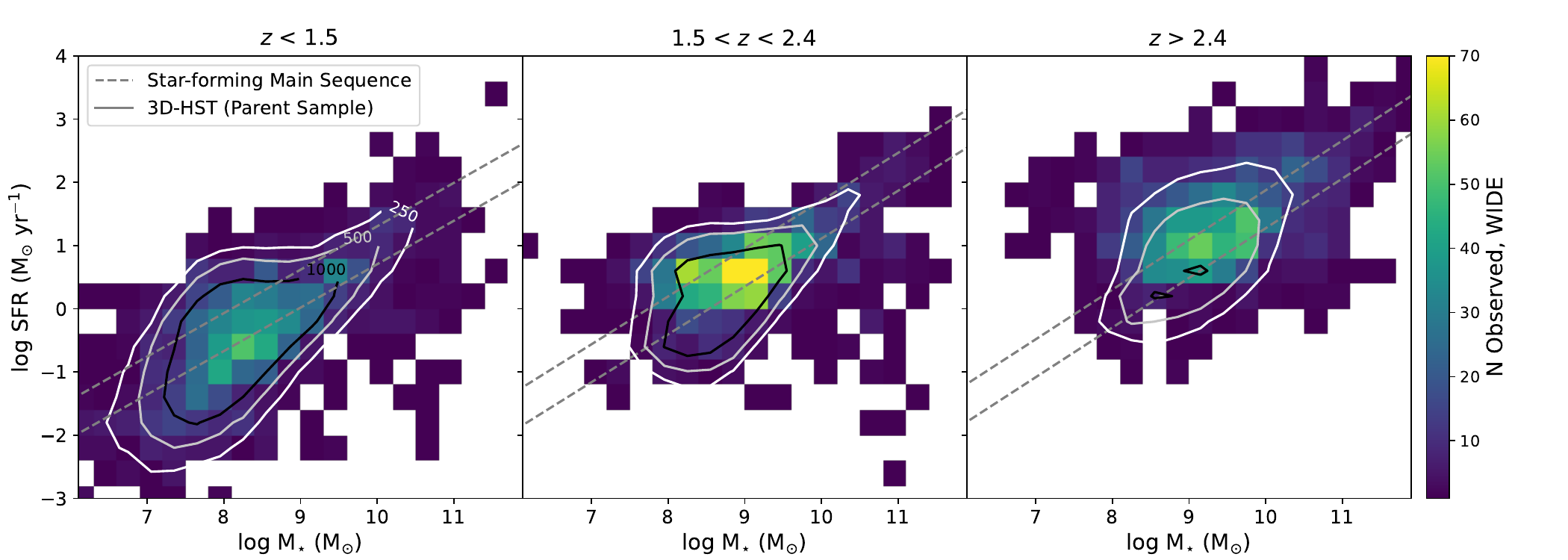}
\end{center}
\caption{Observed galaxies in the Wide survey (colors) compared to the parent catalog from 3D-HST \cite[][contours at 250, 500, and 1000 sources]{2016ApJS..225...27M} in our three primary redshift bins. Across all redshift and mass ranges, Wide is sampling galaxies across the full range of SFRs from 3D-HST, including those that lie below the ``main sequence'' \cite[overplotted at $\pm$1-$\sigma$ as dashed lines at $z=1,2,$ and $3$, respectively;][]{2014ApJS..214...15S}.}
\label{fig:completeness}
  \end{figure*}

  In Figure \ref{fig:completeness}, we show the final completeness of the Wide spectroscopic sample compared to the parent 3D-HST sample in the M$_{\star}-$SFR plane in three different bins of redshift.  Our completeness increases above $z = 1.5$ (Section \ref{sec:pclass}), where we are effectively sampling across the main sequence of star formation.  In the $1.5 < z < 2.4$ range, the distribution of SFR values in WIDE is indistinguishable from the distribution in 3D-HST according to a two-sided Kolmogorov-Smirnov test at masses above 10$^{9.5}$ M$_{\odot}$.
%We adopt a weighting scheme where each subsequent class is worth 
\subsection{Target Acquisition}

Given the small size of each NIRSpec microshutter, care must be taken to ensure a proper alignment between the instrument apertures and the targets is made.  We utilized the HST-based images from the aforementioned CANDELS fields, but with astrometry updated to the Gaia reference frame \citep{2012AA...538A..78L}.

We adopted the MSATA procedure for the actual target acquisition.  This involves observing a number of ``reference objects,'' or compact sources that are within a fixed range of magnitudes so as to be well-detected in a fixed exposure time without being saturated.  The exposure time and the NIRSpec target acquisition filter (F110W, F140X, and CLEAR) can be chosen to maximize the number of observable reference objects, up to 8.  These sources are observed through grids of 13 by 7 open microshutters.  Since the reference objects must be compact and inter-shutter bars occlude a significant fraction of the areal coverage, as well as a number of shutters being completely inoperable, a larger number of usable sources must be constructed for each prospective pointing to ensure a successful acquisition.

We constructed a catalog of usable reference objects again based on the 3D-HST photometric catalog.  Although these catalogs have HST/WFC3 and \textit{Spitzer}/IRAC photometry covering the same wavelength range as the NIRSpec target acquisition filters, care must be taken to accurately predict the exact magnitudes for each object.  We used the \texttt{MAGPHYS} code \citep{2008MNRAS.388.1595D} to fit the broad-band photometry for each object in the 3D-HST catalog.  Based on the best-fit spectral energy distribution for each source, we calculated the predicted magnitude in the NIRSpec target acquisition filters using \texttt{PySynphot} \citep{2013ascl.soft03023S}.  Additionally, we required a source to be spatially compact based on the catalog \texttt{CLASS$\_$STAR} value, typically $\ge$ 0.8 in F160W although in some cases this is relaxed to 0.5 in fields where an insufficient number of reference objects would result.  A final visual inspection of each potential reference object was performed to ensure that e.g. a brighter nearby source would not interfere with the MSATA procedure. 
 When NIRCam imaging is available, we verified that sources with large proper motions were removed by comparing the HST positions (circa 2010) with the NIRCam ones (circa 2023).  The final list of reference objects was added to APT and the optimal setup, typically 7 or 8 reference objects in 3 or 4 MSA quadrants, was used for the target acquisition. This procedure was successfully adopted for all Wide pointings, resulting in a 100 percent target acquisition success rate (with the exception of a single failure due to an observatory guide star issue unrelated to the MSATA procedure).

\subsection{Final Survey Statistics}
\begin{figure*}
\begin{center}
\includegraphics[width=0.95\textwidth]{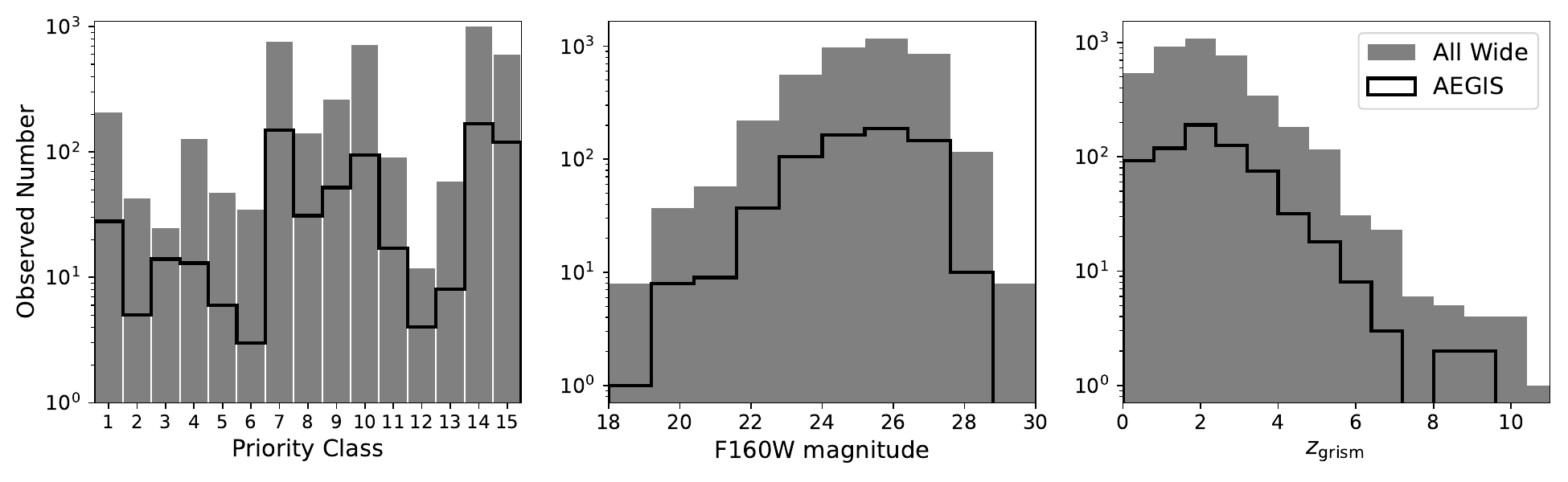}
\end{center}
\caption{Distribution of objects in the full Wide sample (filled histograms) and in the AEGIS sample (open histograms) which makes up the initial Wide data release (Section \ref{sec:release}).  Panels show the distribution in priority classes (Table \ref{tab:pclass}), as well as F160W magnitude and $z_{\mathrm{grism}}$ from 3D-HST \citeauthor{2016ApJS..225...27M}) \citeyear{2016ApJS..225...27M}).}
\label{fig:histograms}
  \end{figure*}

In total, the Wide survey consists of NIRSpec observations of 4,127 unique objects across the five CANDELS fields, including 189 unique ``P1'' sources.  The distribution of their priority classes, F160W magnitudes, and 3D-HST redshifts are shown in Figure \ref{fig:histograms}, including the AEGIS subset which constitutes the first Wide public data release (see Section \ref{sec:release}).

\section{Data Reduction}
\label{sec:red}
We use the same core reduction procedure as other NIRSpec MOS GTO surveys \cite[e.g.][]{2023NatAs...7..622C,2023AA...677A.115C,2023arXiv230611801C,2023arXiv230602467B,2023AA...678A..68S}, developed by the ESA NIRSpec Science Operations Team (SOT) and GTO teams.  This procedure will be described fully by S. Carniani et al. (in prep.), but we summarize here.  Most of the pipeline workflow uses the same algorithms adopted by the official STScI pipeline \citep{2018SPIE10704E..0QA,2022AA...661A..81F}, with small optimizations for our observations.  Namely, we use a finer grid in wavelength with regular steps 
%to improve the sampling of the line spread function 
in the 2D rectification process. We also estimate path losses for each source by taking into account its relative intra-shutter position and assuming a point-like morphology, as in \citet{2023arXiv230602467B,2023arXiv230408516C}: further work will be done in the future to optimize these estimates for extended sources.  The single point of divergence between the Wide reductions and the other GTO reductions are in the sigma-clipping algorithm.  This algorithm does not work well when the number of exposures is low (as in Wide) and does not account for Poisson noise from bright pixels (as is typical for many Wide targets).  Future data releases will utilize an updated sigma-clipping procedure, but the current release (see Section \ref{sec:release}) omits this step. We note that outliers due to cosmic rays and/or hot pixels are still identified during the ramp-to-slope processing step and are removed in the combination process.

The local sky background is subtracted using the nodded exposures (2 nods for G235H and G395H; 3 nods for PRISM).  In the 2 nod configuration, the local sky background is corrected via the direct subtraction of the other exposure.  In the case of 3 nods, it is corrected via the average of the two other exposures in the sequence.  For some very extended objects, the 3-point nodding of the PRISM exposures leads to significant self-subtraction and a 2-point reduction is necessary; see Section \ref{sec:release}. 1D extractions are performed with boxcar apertures.  The default width is 5 pixels corresponding to a single open shutter; 3 pixel and 15 pixel extractions are also available for the 2-point reductions described above.

\section{Science Goals}
\label{sec:science}
The Wide survey covers the largest area of all NIRSpec Cycle 1 programs, more than 320 arcmin$^2$, including all of the other NIRSpec GTO survey tiers (Figure \ref{fig:survey}). Although Wide has considerably less exposure time per spectroscopic pointing than the JADES survey, the large area opens up an observational discovery space that is otherwise unattainable. Specifically, Wide is designed to characterize the galaxy population at intermediate redshifts (here, $1<z<5$) as a complement to the deeper JADES survey reaching to higher redshifts. At the same time, Wide is well suited to study rare objects which would not be observed in great numbers by the other tiers. The main science drivers for the Wide observing strategy are:

\subsection{A census of galaxies at Cosmic Noon}
One of the key design features of Wide is the tiered nature of the priority classes, where more objects are generally available and observed in the lower priority classes.  This leads to a survey with good number statistics in the higher-priority, higher-$z$ categories while also broadly sampling the population of targets with intermediate priorities.  In the case of Wide, this means that 50\% of the observed sample (1909 galaxies; Figure \ref{fig:histograms}) is located at $1.5 < z_{\mathrm{grism}} < 3$, where the cosmic star formation rate density peaks \citep{MadauDickinson}.  This unprecedentedly large sample of galaxies at ``cosmic noon'' will allow us to better understand the buildup of the galaxy population when most of the star formation in the history of the Universe took place.  

\begin{figure*}
\begin{center}
\includegraphics[width=0.95\textwidth]{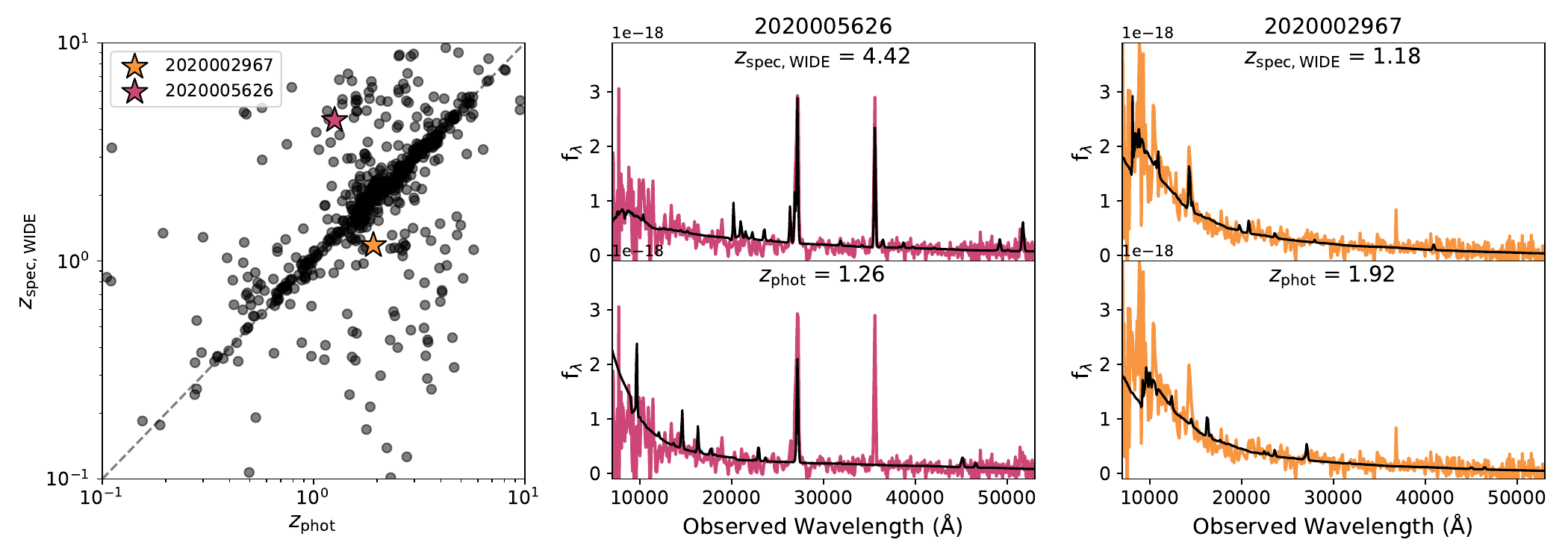}
\end{center}
\caption{(Left) Photometric redshifts (from \texttt{EAzY}) versus Wide spectroscopic redshifts for targets in the AEGIS field constituting the first Wide data release.  While many of the objects had accurate photo-$z$ estimates (70\% have $\mid\Delta$z/(1+$z$)$\mid$ values below 0.25), there is a significant benefit to obtaining precise redshifts from spectroscopy for many of our primary science cases.  (Center and Right) Example PRISM spectra of objects with a discrepancy between $z_{\mathrm{phot}}$ and $z_{\mathrm{spec}}$. 
 The black line shows the best-fitting \texttt{EAzY} template at each of the redshifts, where the Wide-based spectroscopic redshifts are unequivocally correct.}
\label{fig:zphotzspec}
  \end{figure*}

As many of our priority classes explicitly incorporate existing redshift information, an initial key test of the Wide survey will be verification of these redshifts compared to the Wide spectroscopic redshifts (see Section \ref{sec:release} and Figure \ref{fig:zphotzspec}).  Moreover, our relaxed requirements for inclusion into lower priority classes allows Wide to survey any potential population of galaxies that are mis-identified from their photometry alone \citep[e.g. confusion between Lyman-breaks and Balmer-breaks as explained in][]{2023NatAs...7..622C}.

Wide will facilitate the determination of stellar masses from fitting to broadband photometry with these spectroscopic redshifts, star formation rates from \ha\ fluxes, and gas-phase metallicity determinations from strong-line ratios such as \oiii/\hb, \nii/\ha, \oii/\hb, and combinations thereof \cite[e.g.][]{2024AA...681A..70L}. These are the necessary pieces to reconstruct scaling relations such as the ``star formation main sequence'' \cite[the relation between star formation rate and stellar mass;][]{2014ApJS..214...15S} or the stellar mass$-$gas-phase metallicity relation \citep{2023arXiv230408516C} over a large dynamic range in stellar mass and redshift.

\subsection{The evolution of galaxy quenching across cosmic time}
\begin{figure*}
\begin{center}
\includegraphics[width=0.99\textwidth]{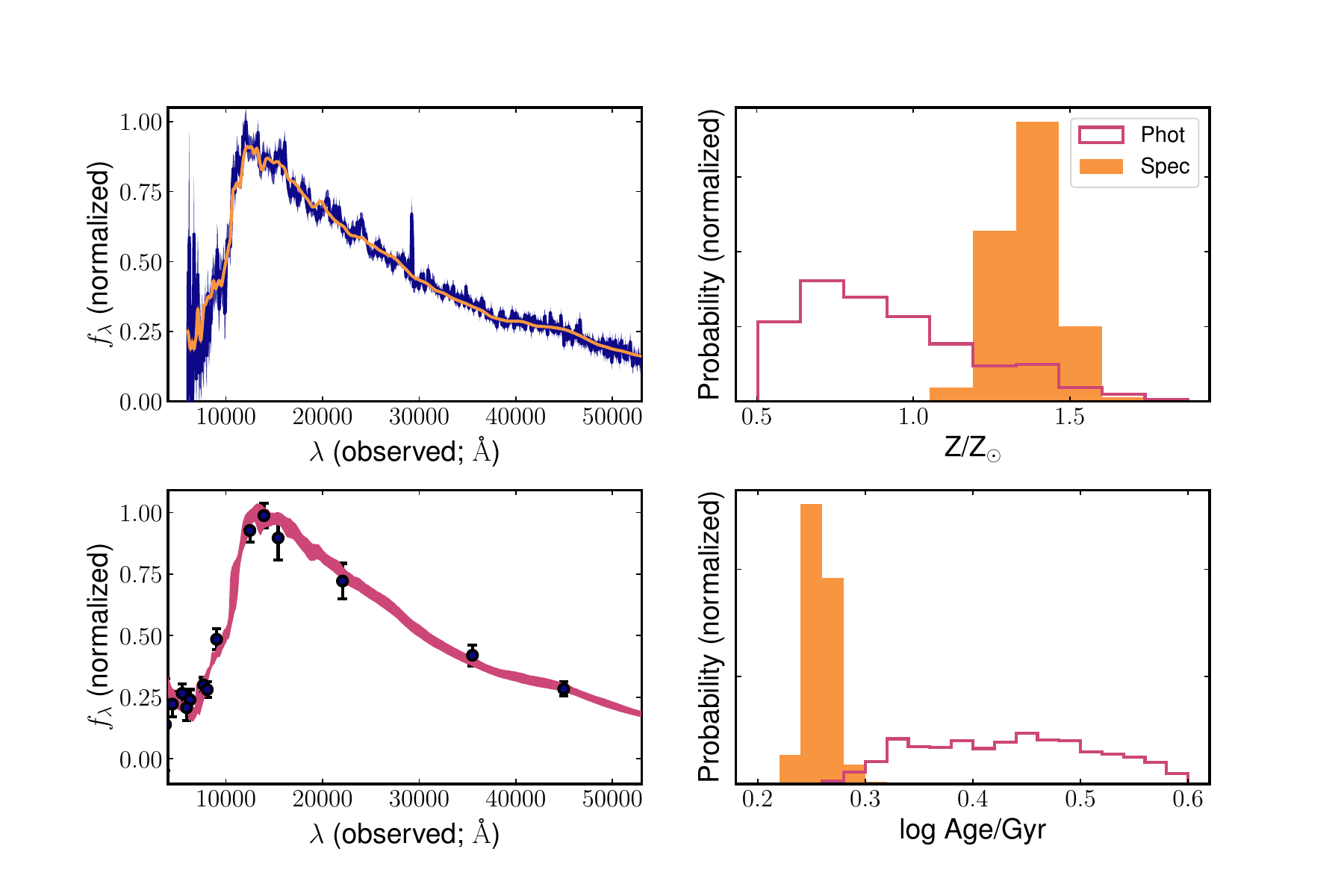}
\end{center}
\caption{Example PRISM spectrum (top left) and photometry (bottom left) for a galaxy in the Wide survey (3D-HST ID: UDS 22634).  Both the spectrum and the photometry are fit with \texttt{Bagpipes} \citep{bagpipes1,bagpipes2}, with the best-fit models and their dispersion overplotted.  The right panels show posterior probability distributions for the stellar metallicity (top right) and age (bottom right) for an exponentially-declining star formation history.  The spectroscopic fit (filled histogram) results in better-constrained posteriors with a significantly different physical interpretation than would be possible with photometry alone \citep[see also][]{2006ApJ...645...44K}.}
\label{fig:specphot}
  \end{figure*}

  The local galaxy population at $M_*>10^{9}$ M$_\sun$ is bimodal with respect to its color, age and star formation rate \citep[e.g.,][]{Kauffmann:2003,Bell:2004,Brinchmann:2004} and hence is split into actively star-forming galaxies and mainly passively evolving systems. The galaxy bimodality is detected out to $z\gtrsim3$ with a significant population of galaxies with no or low-levels of star formation already present at those redshifts \citep{2022arXiv221211638N,2023Natur.619..716C,2023ApJ...949L..23S,2023arXiv230214155L,2023arXiv230805606G,2023arXiv230815011K}. How the galaxy population became bimodal and which mechanisms are mainly responsible for the suppression of star formation remain observationally unsettled questions. With JWST we are able, for the first time, to obtain rest-frame optical continuum spectra out to $z<7$ that allow us to derive star formation histories (SFHs) and stellar population parameters for galaxies across the galaxy population, including direct measurements of the Balmer break from the spectra. Spectral fitting can provide far more information than photometry alone can, and is capable of breaking degeneracies in critical parameters such as age and metallicity (Figure \ref{fig:specphot}). 
  
\begin{figure}
\begin{center}
\includegraphics[width=0.45\textwidth]{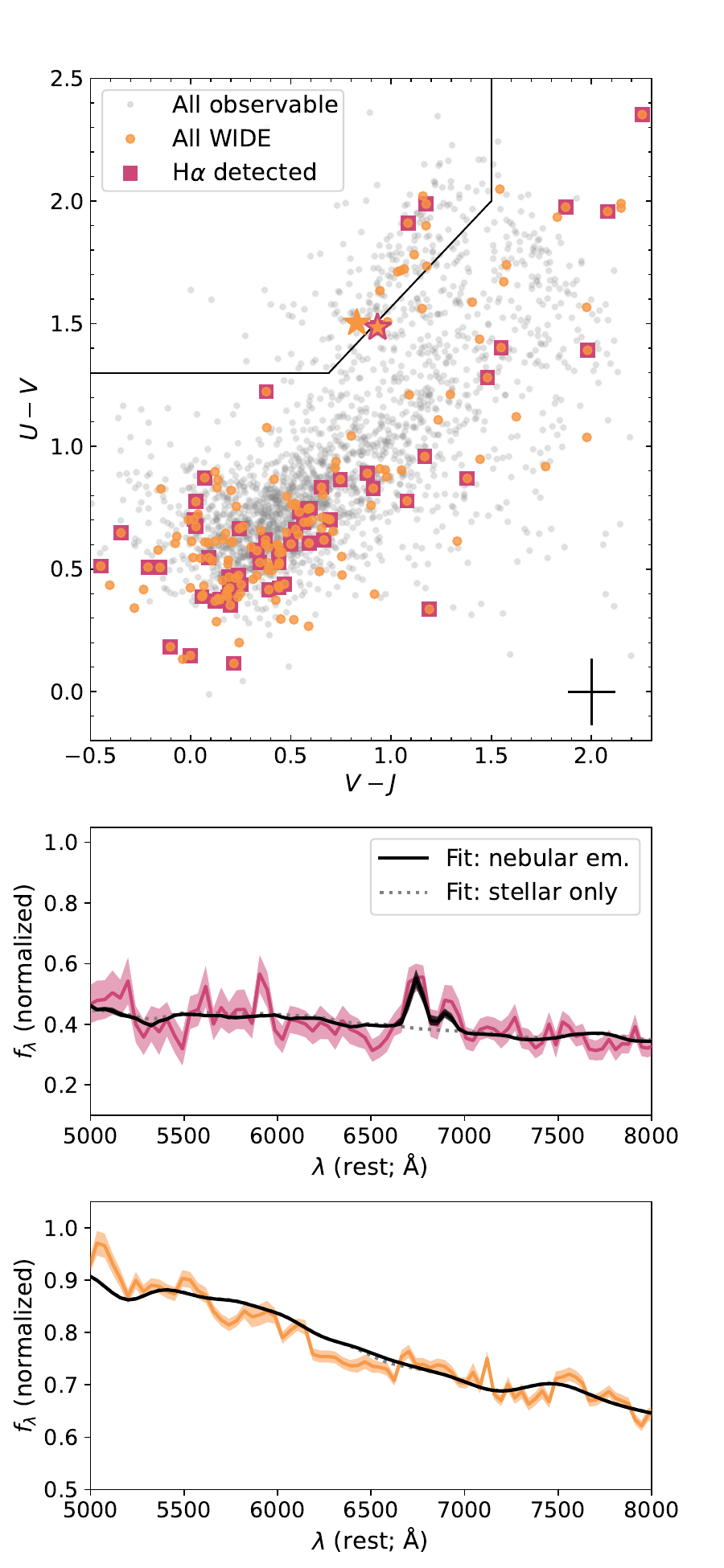}
\end{center}
\caption{Observed UVJ diagram \citep{UVJ}, showing restframe $U-V$ and $V-J$ colors \cite[calculated using EAzY;][]{eazy}, for objects at $1.5 < z < 2.4$ with M$_{\star} > $10$^9$ M$_{\odot}$ in the AEGIS pointings of Wide.  (Upper panel) Solid orange points are the observed sample and transparent grey points are the parent sample.  While objects located within the boundary in the upper-left corner are often considered to be quiescent in nature \cite[e.g.][]{2011ApJ...735...86W,2013ApJS..206....8M}, Wide spectroscopy reveals \ha\ emission in some of them (squares) indicating some source of ionizing photons \cite[e.g. residual star formation, heating by old stellar populations, and/or an AGN;][]{2019AJ....158....2B,2021ApJ...923...18M}.  (Lower panels) Wide PRISM spectra for the two objects denoted by stars in the upper panel; sources close to the boundary between star-forming and quiescent.  While the characteristic uncertainty in the restframe colors is large (shown in the lower-right corner of the upper panel), spectroscopy can unambiguously differentiate via e.g. a detection of \ha\ emission at 6563 \AA\ in excess of the \texttt{BAGPIPES}-determined stellar continuum.}
\label{fig:uvj}
  \end{figure}

 Moreover, while galaxies can generally be separated into star-forming and quiescent on the basis of their restframe UVJ colors \citep[e.g.][]{UVJ}, spectroscopy can reveal the presence of emission lines like \ha, which could be an indication of AGN activity, hidden star formation, or the presence of hot, evolved stars in an otherwise passive galaxy \cite[e.g. Figure \ref{fig:uvj};][]{2019AJ....158....2B}. With spectroscopic data, reliable and consistent methods can be used to track the evolution in star formation rates (SFRs) from the local Universe out to high redshifts. The NIRSpec Wide survey complements the deeper JADES tiers by focusing on redder and more massive galaxies, providing large number statistics for the full population of galaxies in the intermediate redshift range $1<z<5$ connecting the present day and infant Universe. 

\subsection{Rest-frame optical spectra of \textit{Spitzer}/IRAC excess sources}
  Prior to NIRSpec, most spectroscopic observations of the highest-redshift sources were restricted to the rest-frame UV (observed NIR).  However, techniques have been developed using broadband \textit{Spitzer}/IRAC photometry as ``low-resolution'' spectroscopy over large areas: Lyman-break galaxies are selected via optical/NIR imaging, where optical spectral features are contained inside the IRAC bandpasses in certain redshift windows.  Strong photometric excesses in these bands were attributed to emission lines \citep[e.g. \ha\ and/or \oiii;][]{2013ApJ...763..129S,Smit:2015,Roberts-Borsani:2016} or to a strong Balmer break \citep[e.g.][]{2005MNRAS.364..443E,Hashimoto:2018}.  Bright galaxies with the strongest photometric excesses are sufficiently rare ($\sim$50 across all five CANDELS fields) so that Wide is the only GTO tier that can realistically probe the full population. Low-resolution spectra with NIRSpec trivially measure the bright emission lines (if present) for accurate redshifts, metallicities, and SFR measurements.  These galaxies are the brightest sources available at $z\sim6-8$ and are well suited for more detailed follow-up studies.

\subsection{Ionized gas outflows from SF and AGN galaxies at $2<z<5$}
  Massive gas outflows from galaxies are thought to be an important mechanism to regulate star formation in galaxies \citep[e.g][]{Silk:1998,King:2003,Hopkins:2008a} and to enrich the inter-galactic medium with metals. Galactic outflows have often been detected as broad wings, typically blue-shifted, of both forbidden and permitted emission lines \citep[e.g.][]{Mullaney:2013,Harrison:2016,Zakamska:2016}. Ionized gas outflows are prevalent in starburst galaxies \citep[e.g.][]{Westmoquette:2012,Cicone:2016}, but are particularly strong in AGN host galaxies \citep[e.g.][]{Arribas:2014,Carniani:2016,Husemann:2019a}. While systematic analyses of ionized gas outflows from rest-frame optical emission lines have been performed from the ground already up to redshift $z\sim2.5$ \citep{2012ApJ...761...43N,2019ApJ...875...21F}, only JWST is able to probe the redshift range beyond $z>2$ based on H$\alpha$ \citep{2023arXiv230611801C,Xu2023}. The detection of the outflows and their characterization requires a high spectral resolution to properly resolve the line shapes into multiple components.% (see Figure \ref{fig:outflow}).

%    \begin{figure*}
%\begin{center}
%\includegraphics[width=0.9\textwidth]{spec2893.pdf}
%\end{center}
%\caption{Example R2700 G235H/F170LP spectrum for a galaxy in the Wide survey (3D-HST ID: AEGIS 23068; $z=2.74$).  There is a clear asymmetry to the \oiii\ and \hb\ emission lines indicative of an outflow.  This outflow can be fit by adding an additional broad component that is blueshifted by 236 km s$^{-1}$ compared to the systemic velocity of the galaxy.  For both individual objects as well as stacks, the Wide survey will probe galactic outflows at redshifts that are difficult or impossible to do with ground-based instrumentation.}
%\label{fig:outflow}
%  \end{figure*}

\subsection{The evolution of black hole mass -- host galaxy scaling relations}
  The tight relation between black hole (BH) mass and host galaxy properties, such as stellar velocity dispersion $\sigma_*$ and bulge mass $M_\mathrm{bulge}$, have been established as a fundamental characteristic of massive galaxies \citep[e.g.][]{Gebhardt:2000,Ferrarese:2000,2004ApJ...604L..89H}. Initially, BH and galaxy growth were assumed to be regulated via AGN feedback, but it has been shown that simply the merging of galaxies and their BHs naturally lead to tight BH mass -- host galaxy mass relation due to the central limit theorem \citep{Peng:2007,Jahnke:2010}. It is vital to explore the evolution of the relation between BH mass and their host galaxies to further understand its origin and discriminate between different pathways. However, it is challenging to probe the BH mass -- host galaxy scaling relations beyond $z>1$ \citep{Jahnke:2009,Merloni:2010}, and these studies have reported inconsistent results. With JWST NIRSpec spectroscopy we aim to measure $\sigma_*$ from the redshifted Ca triplet in the stellar continuum of unobscured AGN at $1<z<2$, or estimate $\sigma_*$ from narrow emission lines. %In addition, we can explore the BH mass -- gas phase metallicity relation \citep[e.g.][]{Matsuoka:2011} based on optical line diagnostics as a tracer for the stellar mass.  
  In addition, we can explore the BH mass -- stellar mass relation by performing joint stellar population and AGN modeling to the PRISM spectroscopy \citep[e.g.][]{2023arXiv230801230M,2024arXiv240302304W}.  
  This opens the possibility to study the evolution in the BH mass -- host galaxy relation out to $z\approx6$ with the Wide survey, extending JWST results to much larger samples and a wider range in black hole mass than are currently available \cite[e.g.][]{Greene2023,2023arXiv230801230M,2023AA...677A.145U}. 
       
\subsection{The dynamics of galaxies including dark matter content}
  The kinematic structure of galaxies is observed to evolve over time, as ionized gas velocity dispersions of galaxies increase with redshift, while galaxy sizes and rotational support decrease \citep[e.g.,][]{Wel:2014,Wisnioski:2015, Simons:2016,Uebler:2019}. Theoretical studies show that star-forming galaxies at high redshift ($z\sim2$) may consist of gas-rich turbulent disks, which grow through cold accretion and minor mergers \citep[e.g.][]{Dekel:2009, Ceverino:2012, Genel:2012}. Measurements of the kinematics of high-redshift galaxies therefore provide crucial insight into galaxy mass assembly and feedback mechanisms.
  Beyond $z\gtrsim2$, however, the rest-frame optical emission lines typically studied become extremely challenging to observe. Although much progress at $1<z<3$ has been made in recent years with ground-based surveys to infer the internal motions of typically massive and extended galaxies using the KMOS, MOSFIRE and SINFONI near-IR spectrographs \citep{Kriek:2015, Wisnioski:2015, Simons:2016, 2016MNRAS.457.1888S,Turner:2017, Foerster-Schreiber:2018,Price2020}, and using ALMA observations of rest-frame FIR lines in bright galaxies at $z\sim4-7$ \cite[e.g.][]{Smit:2018,Neeleman2020,Jones2021,Lelli2021,Rizzo2021}, the precise physical mechanisms that govern galaxy kinematics remain unclear, particularly for low-mass and compact systems. Observational limitations, including ground-based seeing and OH skyline contamination, are a major motivator to undertake these investigations with NIRSpec.  
  
  With the NIRSpec Wide survey, we will systematically explore the internal galaxy kinematics based on the widths of integrated rest-frame optical emission lines out to high redshifts and even study rotation curves from spatially-resolved kinematics for a significant subset of galaxies. Through dynamical modeling of the kinematics, we will constrain the total mass budgets of the galaxies, including dark matter on galactic scales \citep{deGraaff2023}. 
 As Wide is targeting several thousand galaxies spanning a wide range in both stellar mass and redshift, we will be able to trace the evolution in the galaxy mass budget and explore the physical origins and evolution of galaxy kinematics.

\subsection{Galaxy Clustering Estimates at $z\approx2-6$}
In the $\Lambda$CDM framework, galaxies are formed in the centers of dark matter halos \cite[e.g.][]{1978MNRAS.183..341W,1984Natur.311..517B,1991ApJ...379...52W}, hence their properties are closely linked to the properties of the halos they reside in. Assessing this galaxy-dark matter relation gives insight into the efficiency of stellar mass assembly and places key constraints on galaxy formation and evolution models. A commonly used method to assess the galaxy-dark matter connection and provide a diagnostic for distinguishing theoretical models is measuring the clustering of galaxy populations using pairwise velocity dispersion and two-point correlation function methods \citep{1980lssu.book.....P,1983ApJ...267..465D}. Studies of $\sim$10-100s of thousands of galaxies have linked galaxies to their local environment and indicated only a slight evolution of galaxy clustering across $z\approx0-2$, with the strength depending on galaxy type and physical properties \cite[e.g.][]{2006MNRAS.368...37L,2011ApJ...736...59Z,2014AA...568A..24B,2018MNRAS.474.3435L}. There has been an effort to extend the analysis to higher redshift \cite[$z=2-6$; e.g.][]{2017ApJ...841....8I,2018MNRAS.477.3760H,2018PASJ...70S..11H}, however, sample statistics and/or uncertainties associated with photometric surveys make it difficult to put stringent constraints on the stellar-halo mass relation. Nevertheless, spectroscopic studies \cite[e.g.][]{2015AA...583A.128D} have suggested that the stellar mass-halo mass ratio at $z\approx$ 3 might be higher than expected from theoretical models. Larger spectroscopically-confirmed samples extending to higher redshifts are required to assess galaxy clustering at $z\gtrsim$ 2, with JWST expected to deliver the required sensitivity out to $z\approx$ 10 \citep{2020MNRAS.493.1178E}. NIRSpec Wide will provide the accuracy and precision ($\lesssim$ 100 km s$^{-1}$ for S/N $\gtrsim$ 5 in G235H/G395H) over a large volume to identify galaxy pairs across $z\approx2-6$, with $\sim$2500 expected galaxy pairs at $z\gtrsim$ 2.5. This survey will allow us to assess the pairwise velocity dispersion and derive the two-point correlation function for subsets of the sources across $z\approx2-6$ to infer the dark matter halo masses and compare with simulations and observational clustering measurements of the expected descendants in the local Universe.

\subsection{Spatially-resolved spectroscopy of galaxies at Cosmic Noon}
\label{sec:extended}

\begin{figure*}
\begin{center}
\includegraphics[width=0.95\textwidth]{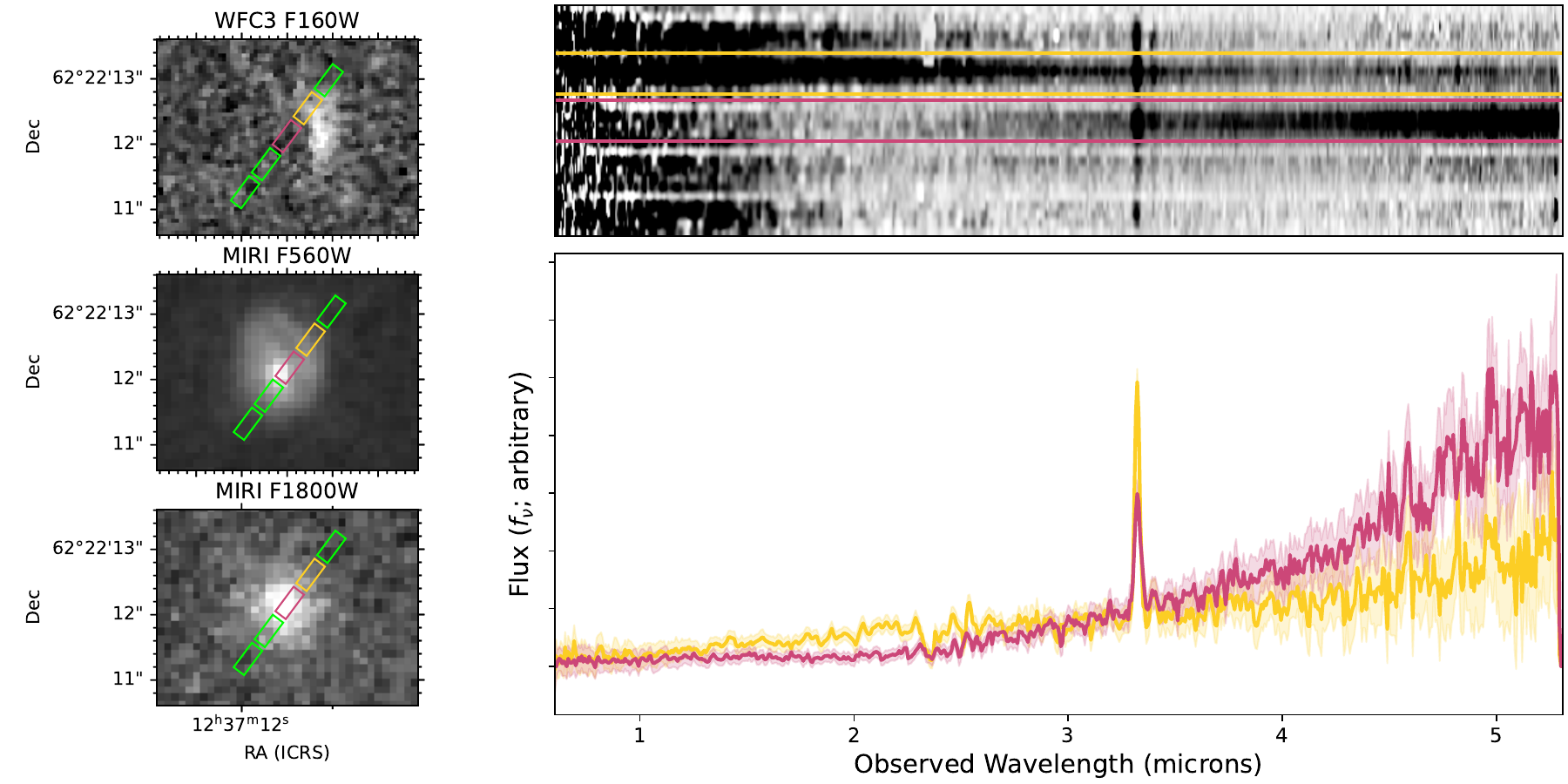}
\end{center}
\caption{Example Wide data for GN20 \citep{2006MNRAS.370.1185P,2023arXiv231203074B,2024arXiv240218672C}, a bright submillimeter galaxy in the GOODS-N field at $z=4.05$.  (Left) HST/WFC3 and JWST/MIRI images of the system with the NIRSpec microshutters overplotted (the 1$\times$3 configuration adopted in Wide along with the $\pm$1 shutter above and below observed in the two different nodding positions).  As noted in \citet{2015ApJ...798L..18H} and \citet{2023arXiv231203074B}, there is significant dust attenuation in the center of the system along with an unobscured (UV-bright) component visible in e.g. F160W.  (Right) The 2D and 1D Wide spectra for this object, with extractions peformed for a single shutter (colors).  The unobscured component (upper shutter) is well-detected in \ha, while the obscured component is observed with a prominent red continuum.}
\label{fig:GN20}
  \end{figure*}

  Recent JWST/NIRSpec integral field spectroscopic (IFS) studies of galaxies at early cosmic times (i.e. $3 < z < 9$) have revealed the frequent presence of clumps and companions \citep{2023arXiv230914431R,2024AA...682A.122J}, metallicity gradients \citep{2023arXiv231200899A}, outflow-induced gas turbulence \citep{2023arXiv230905713P}, off-centered and dual AGN \citep{2023arXiv231203589U,2023AA...679A..89P,2024AA...681C...2P}, and other complex structures that are difficult to interpret without spatially-resolved spectroscopy.  Although limited along the cross-dispersion direction, Wide offers the possibility to probe spectrally the spatial structure of individual objects (Figure \ref{fig:GN20}).  Additionally, Wide will obtain average radial profiles of key properties (e.g. metallicity, ionization, velocity dispersion) for large samples, potentially an order of magnitude higher than with the GTO GA-NIFS survey.  This information is of paramount importance to compare with the predictions of current high-resolution cosmological zoom-in simulations of high-$z$ galaxy populations \cite[e.g.][]{2016MNRAS.460.2731C,2021MNRAS.504.4472C,2024arXiv240208911N}.

\section{Data Release and Plans}
\label{sec:release}

The full suite of reduced data products for the Wide pointings in the AEGIS field (Program 1213) is available on MAST: \url{https://archive.stsci.edu/hlsp/wide}. The target catalog can be downloaded at \url{https://keeper.mpdl.mpg.de/d/9c032a159ffe4cea9160/}. The distribution of their priority classes, F160W magnitudes, and 3D-HST redshifts compared to the full Wide sample are shown in Figure \ref{fig:histograms}. Additionally, we commit to similar releases for the remaining four fields in the future.

A key contribution of the Wide survey is spectroscopic confirmation of the redshifts for the target sample.  For the five Wide AEGIS pointings, our initial redshift catalog based on the PRISM data (forthcoming with the data release described here) results in 435 secure redshifts from 713 total targets, resulting in a 61\% success rate.  The remaining targets are predominantly in lower priority classes: 272 of the 278 without Wide spectroscopic redshifts are in priority class 7 or below, 196 are in priority class 10 or below, and 148 are in priority class 14 or below.  This is comparable to the results from the JADES Deep survey \citep{2023arXiv230602467B}, which obtained a 70\% spectroscopic redshift confirmation rate with similarly lower success rates for lower priority classes.

\begin{figure*}
\begin{center}
\includegraphics[width=0.95\textwidth]{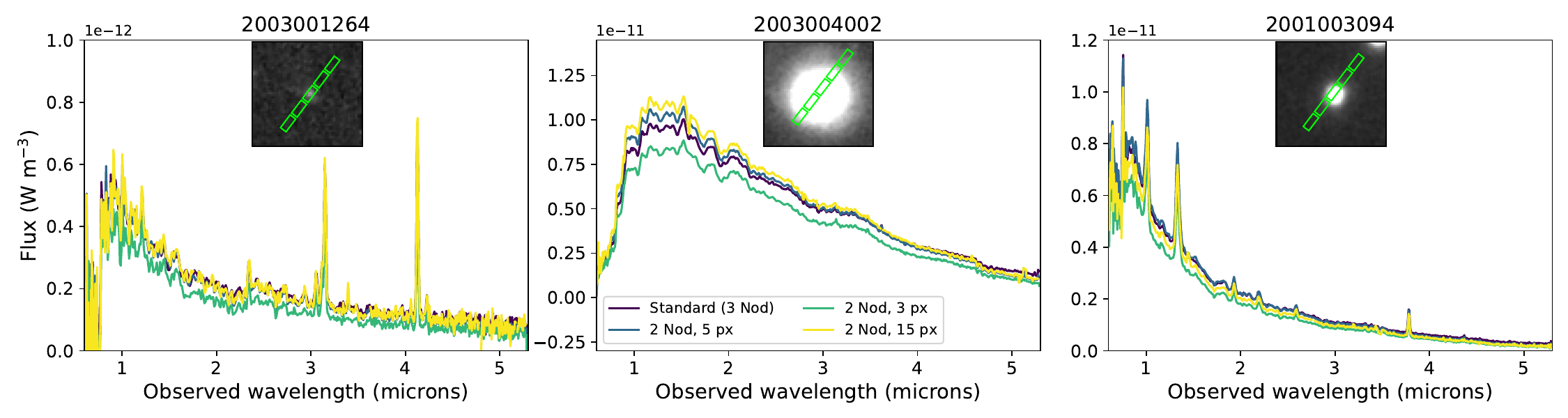}
\includegraphics[width=0.95\textwidth]{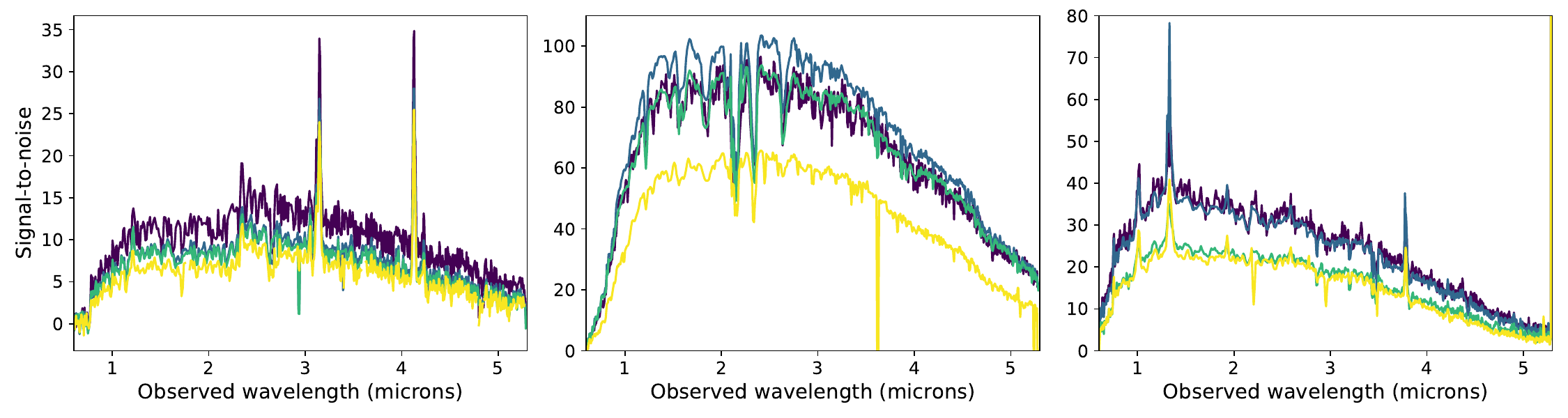}

\end{center}
\caption{Example Wide PRISM spectra (top) and signal-to-noise (bottom) for three objects in GOODS-N, showing the effect of the choice of the nodding pattern and extraction on the resulting 1D spectra.  Insets show the HST/WFC3 F160W images with the NIRSpec MSA shutters overplotted; note a 1$\times$3 microshutter configuration with the observed three-point nodding in the PRISM results in a projected area of 1$\times$5 shutters on the sky.  (Left) A compact object, well-centered in the MSA: there is very little difference between the different nodding and extraction choices.  (Center) A very extended object: in this case, self-subtraction becomes an issue with the three-point nodding resulting in a higher signal and signal-to-noise when the two-point reduction is used. (Right) An intermediate case: the choice of different extractions can be motivated scientifically (see e.g. Section \ref{sec:extended}).}
\label{fig:reduction}
  \end{figure*}

 This release includes reduced 1D and 2D spectra, including multiple options for the data processing and 1D spectral extractions, as well as a catalog of redshifts determined from the spectroscopy (see Figure \ref{fig:zphotzspec}).  Specifically, we process the PRISM data in two different ways: (1) in an identical fashion to the grating data, taking the sum of the exposures in the top and bottom microshutters only (``2 nods''), as well as (2) using all three nods (``3 nods'').  The former is particularly useful when a source is very extended spatially and hence background self-subtraction from closely-nodded exposures can cause issues (see the central panel in Figure \ref{fig:reduction}).  In the 2 nods case, we also include three different spectral extraction sizes: 3, 5 (default), and 15 pixels in the cross-dispersion direction. While the larger extractions include more flux, which is useful in the case of an extended object, it increases the chance of contamination from hot pixels and/or cosmic ray hits as well as increases the total amount of detector noise in the extracted spectrum. Care should be taken when deciding which extraction and nodding pattern to use for which type of object: see Figure \ref{fig:reduction}.  We also note that the same point-source slit loss correction is applied to all these different 1D and 2D extractions, and therefore caution that additional flux calibration is needed for sources that are resolved by JWST/NIRSpec ($ \gtrsim 0\farcs1$).

\section{Conclusions and Outlook}

The NIRSpec GTO Wide program is an ambitious Cycle 1 project designed to study the evolution of the galaxy population at $1 < z < 5$, with extensions to higher redshifts for some bright, rare targets.  Wide was conceived to be the optimal way to rapidly tile the sky to maximize the observability of high-value targets with a spatial density of a few per MSA field-of-view while simultaneously providing a much larger census of galaxy spectra with a wavelength coverage and sensitivity that is unmatched by other facilities.  The 105 hours allocated to the program result in 31 pointings, spread across the five CANDELS fields with ancillary photometric information from HST. 

The science cases for Wide are varied (Section \ref{sec:science}), including a large census of galaxies at cosmic noon, the assembly of star-forming galaxies in the early Universe,  the incidence rate of galaxy-scale outflows across a large dynamic range in stellar mass, and the nature of accreting supermassive black holes and their relationship with their host galaxies.  Spectroscopy can provide additional information that cannot be obtained by photometry alone. These science cases naturally lead to an observing strategy like the one adopted here, combining low- and high-resolution spectroscopy for a large number of targets spread across the sky.  Wide accomplishes these and other science goals with a unique target selection facilitated by the eMPT software and a results-oriented prioritization strategy.
  
Future data releases from Wide will provide a tremendous legacy value to the community, representing a detailed census of galaxies and AGN focused on the epoch of Cosmic Noon.  The observing strategies, prioritization schemes, and data reduction techniques will also serve as a blueprint for future large spectroscopic surveys with NIRSpec.  And, together with its partner JADES and the GA-NIFS programs, these surveys will form the core of spectroscopic information about galaxies in the early Universe for years to come.

\begin{acknowledgements}
The authors would like to thank Bernd Husemann for critical contributions to the design and implementation of the Wide survey.  The authors would also like to thank Pierre Ferruit and Peter Jakobsen for technical and organizational support throughout the development of the survey, and for their stewardship of the NIRSpec instrument and team over the years.
The NIRSpec MOS GTO team thanks the Instrument Development Teams and the instrument teams at the European Space Agency and the Space Telescope Science
Institute for the support that made this program possible. We also thank our program coordinators and staff at STScI, including Andrew Fox, Diane Karakla, Crystal Mannfolk, Blair Porterfield, and Glenn Wahlgren for their help in planning our observations. MVM is supported by the National Science
Foundation via AAG grant 2205519, the Wisconsin Alumni Research Foundation via grant MSN251397, and NASA via STScI grant JWST-GO-4426.
SC acknowledges support by the European Union's HE ERC Starting Grant No. 101040227 - WINGS.
IL is supported by the National Science Foundation Graduate Research Fellowship under Grant No. 2137424.
SA acknowledges support from Grant PID2021-127718NB-I00 funded by the Spanish Ministry of Science and Innovation/State Agency of Research (MICIN/AEI/ 10.13039/501100011033). 
AJB, AJC, AS \& GCJ acknowledge funding from the ``FirstGalaxies'' Advanced Grant from the European Research Council (ERC) under the European Union’s Horizon 2020 research and innovation programme (Grant agreement No. 789056).
RM, FDE, and JW acknowledges support from the Science and Technology Facilities Council (STFC), by the ERC through Advanced Grant 695671 ``QUENCH'', by the UKRI Frontier Research grant RISEandFALL.
H{\"U} gratefully acknowledges support by the Isaac Newton Trust and by the Kavli Foundation through a Newton-Kavli Junior Fellowship. 
RS acknowledges support from a STFC Ernest Rutherford Fellowship (ST/S004831/1). 

\end{acknowledgements}

\bibliographystyle{aa}
\bibliography{wide}

\end{document}